\documentclass[draft]{agujournal2019}
\usepackage{url} 
\usepackage{lineno}
\usepackage[inline]{trackchanges} 
\usepackage{soul}

\draftfalse

\journalname{Geophysical Research Letters}

\begin{document}

\title{Observational Properties of Harmonic EMIC waves: Statistical Study}

\authors{Shujie Gu\affil{1}, Xu Liu\affil{1}, Lunjin Chen\affil{1}, M. E. Usanova\affil{2}, Wenyao Gu\affil{1}, Zhiyang Xia\affil{1} }

\affiliation{1}{William B. Hanson Center for Space Sciences, Department of Physics, University of Texas at Dallas, Richardson, Texas, USA}
\affiliation{2}{Laboratory for Atmospheric and Space Physics, University of Colorado Boulder, Boulder, CO, USA}

\correspondingauthor{Shujie Gu and Xu Liu}{shujie.gu@utdallas.edu;xu.liu1@utdallas.edu}

\begin{keypoints}
\item Harmonic EMIC waves concentrate outside the dayside plasmasphere at $ L >5 $.
\item Harmonic EMIC waves are associated with a low $f_{pe}/f_{ce}$ and a high proton $\beta_H$. 
\item Most of the harmonic EMIC waves are accompanied by a strong EMIC fundamental mode.

\end{keypoints}

\begin{abstract}
Electromagnetic ion cyclotron (EMIC) waves are discrete electromagnetic emissions separated by multiple ion gyrofrequencies. 
Harmonic EMIC waves are defined as waves with a strong electric or magnetic field (or both) at the harmonics of the fundamental EMIC mode. 
In this paper, for the first time, we present a statistical study on harmonic EMIC waves by the Van Allen Probes. 
The EMIC waves are categorized into three types based on their harmonics: (1) fundamental mode only (without higher harmonics), (2) electrostatic (ES) harmonics, and (3) electromagnetic (EM) harmonics. 
Our statistical study shows that ES and EM harmonic EMIC waves predominantly occur on the dayside, outside the plasmasphere with $L >5$ and are associated with a low $f_{pe}/f_{ce}$, a high proton $\beta_H$, and a strong fundamental EMIC mode. 
The results will advance our understanding of harmonic EMIC waves and their generation mechanisms. 
	

\end{abstract}

\section*{Plain Language Summary}
Electromagnetic ion cyclotron (EMIC) waves play an essential role in the acceleration and precipitation of charged particles in the Earth's magnetosphere.
They can resonate with MeV electrons and keV ions in the Earth's magnetosphere and produce the rapid loss of radiation belt electrons and ring current ions.
EMIC waves are sometimes accompanied by electrostatic or electromagnetic harmonic structures in their frequency spectrum, though the properties of these harmonics remain poorly understood. 
In this paper, we present the first statistical study of harmonic EMIC waves with a narrow-band fundamental mode, as observed by the Van Allen Probes. 
Based on their harmonic structures, we categorize the EMIC wave events into three types: fundamental mode only (without harmonics), electrostatic (ES) harmonics, and electromagnetic (EM) harmonics. 
Our statistical study shows that ES and EM harmonic EMIC waves mainly concentrate on the dayside outside the plasmasphere with $L >5$ and are associated with a low $f_{pe}/f_{ce}$, a high proton $\beta_H$, and a strong fundamental mode. 
The statistical findings from these observations provide a foundation for further exploration of the mechanisms behind harmonic EMIC wave excitation.

\section{Introduction}

Electromagnetic ion cyclotron (EMIC) waves are electromagnetic emissions that occur within the Pc1-Pc2 geomagnetic pulsation frequency range, typically between 0.1 and 5 Hz \cite{Allan_1992, Hartinger_2022fAS, Cornwall_1965JGR}. 
They have been widely observed in the Earth magnetosphere \cite<e.g.,>{Anderson_1990GRL}, solar wind \cite<e.g.,>{Jian_2010JGR}, and other planets in the solar system \cite<e.g.,>{Barbosa_1993JGR, Boardsen_2012JGR}. 
Typical EMIC waves propagate in a quasi-parallel direction and exhibit left-hand polarization below ion cyclotron frequencies. In the presence of heavy ions, such as helium (He$^+$) and oxygen (O$^+$), magnetospheric EMIC waves can be divided into three bands: hydrogen, helium and oxygen bands, respectively \cite{Koskinen_2022}. 
They can be excited by a pitch-angle anisotropy in energetic ions with energies of a few tens of keV \cite{Gendrin_1984JGR, Jordanova_2006JGR}, often caused by the injection of plasma sheet particles during geomagnetic storms \cite<e.g.,>{ChenLJ_2009JGR, ChenLJ_2010JGR_emic}, or by magnetospheric compression due to increased solar wind pressure or interplanetary shocks \cite<e.g.,>{Tsurutani_2016JGR, YinZ_2022JGR, Usanova_2010JGR}. 
EMIC waves preferentially occur in the afternoon sector near the magnetic equator \cite{Allen_2015JGR} and on the dayside at large $L$-shells \cite{Usanova_2012JGR}. They become oblique along their propagation paths and undergo polarization reversal as the wave frequency crosses the crossover frequency at high latitudes \cite{Fraser_2001JASTP, LiuYH_2013JGR}.

EMIC waves play a crucial role in the acceleration and precipitation of charged particles in the inner magnetosphere. 
They can resonate with relativistic ($\sim$MeV) electrons through the gyroresonance, leading to the rapid loss of radiation belt electrons \cite<e.g.,>{Thorne_1971JGR, LiuKJ_2012JGR, Blum_2015GRL, Jordanova_2008JGR}. 
Electrons with lower energies (tens to hundreds keV), can interact with EMIC waves via the bounce resonance, resulting in significant pitch angle scattering \cite{Blum_2019JGR, CaoX_2017JGR}. 
In addition, EMIC waves can interact with ring current ions through gyroresonance and cause efficient ion precipitation \cite{Jordanova_2001JGR, Jordanova_2006JGR}. Recent studies have shown that EMIC waves can deplete radiation belt electrons and ring current ions simultaneously  \cite<e.g.,>{Usanova_2014GRL, Lyu_2022GRL}.

Recently, harmonic structures of EMIC waves have been reported. \citeA{ZhuH_2019GRL} observed EMIC waves exhibiting strong electrostatic spectra at multiple and fractional frequencies of the fundamental EMIC waves. 
They proposed that these electrostatic harmonics are generated by wave-wave resonance between fundamental EMIC waves and an accompanying density mode, a density fluctuation caused by the compressional component of the wave's associated electric field. This mechanism was verified by using the cross-bicoherence of low-frequency EMIC waves, the density mode, and high-frequency EMIC waves.
Moreover, \citeA{DengD_2022GRL} observed harmonic EMIC waves with spectral peaks in both electric and magnetic fields, suggesting that higher-order harmonics are excited by wave-wave couplings between lower-order waves. They emphasized the key role of the second harmonic in this process, which has been supported by hybrid simulations and theoretical analysis \cite{XueZ_2022bGRL, YuX_2021GRL}.


Despite the case studies mentioned above, a comprehensive statistical analysis of harmonic EMIC waves remains lacking. In this paper, we utilize the Van Allen Probes data to conduct a statistical survey of harmonic EMIC waves characterized by the existence of electrostatic or electromagnetic harmonics and investigate their associated plasma conditions.
Section 2 briefly introduces the Van Allen Probes data set and data methodology. 
The statistical results are presented in Section 3, followed by conclusions and discussions in Section 4.

\section{Data Set and Methodology}
We use data of the two Van Allen Probes from 2012 to 2019 to perform the statistical study. 
Magnetic field measurements are obtained from the Level-3 64 samples/s data in the Geocentric Solar Ecliptic (GSE) coordinate system, recorded by the fluxgate magnetometer sensor, part of the Electric and Magnetic Field Instrument Suite and Integrated Science (EMFISIS) suite \cite{Kletzing_2013}. 
Electric field measurements are obtained from the Level-2 32 samples/s electric field data in the satellite's UVW coordinates, provided by the Electric Field and Waves (EFW) instrument \cite{Wygant_2013}. 
The electric field measurement $u$ and $v$ components (in the spin plane) are used, and the $w$ component along the spin axis is abandoned due to its high inaccuracy. 
We transform the magnetic fields into a field-aligned coordinate (FAC) system.
The background magnetic field $\mathbf{B_0}$ is calculated based on the time-averaged magnetic field data with a sliding window of 101 points, nearly 1.5 seconds. 
To calculate the electric and magnetic field spectra, we apply the fast Fourier transform (FFT) with a sliding window size of 1024-point time series and a 60-point overlap. 
The wave normal angles (WNAs) are determined through the singular value decomposition method \cite{Santolik_2003} based on the magnetic field.

Plasma density data are acquired from the Level-4 data, derived from the upper hybrid resonance frequency line identified from the electric field power spectrum by EMFISIS \cite{Kurth_2015JGR}. The plasmapause is treated as a boundary layer with a finite width. Thus, the inner magnetosphere is divided into three regions: inside the plasmasphere, outside the plasmasphere, and at the plasmasphere boundary layer (PBL). 
The PBL is determined based on the criteria in \citeA{ LiuX_2020GRL} and \citeA{GuW_2022JGR}.
Their method ensures that the PBLs are associated with sharp density gradients and pronounced density drops over $L$-shell, and only one PBL exists in each inbound or outbound orbit. 
Ion pitch-angle distributions are derived from the differential flux measured by the Helium, Oxygen, Proton, and Electron (HOPE) Mass Spectrometer \cite{Funsten_2013, Spence_2013}. 
The ion temperature in an arbitrary direction is calculated as $ T={1\over n} \int f_i(\textbf{v}) m_i v^2 d^3v$, where $n$ represents number density and $f(\textbf{v})$ is the phase space density in velocity space.
The background plasma beta $\beta$ is defined as the ratio of the plasma thermal pressure ($p=nk_B T$) to the magnetic pressure ($p_{mag}=B^2/2\mu_0$).

\section{Results}
We survey data from the Van Allen Probes throughout its mission (from 2012 to 2019) to identify EMIC wave events. We manually select events with narrow-band frequency, lasting at least five minutes, where the wave frequency remains nearly constant relative to the ion gyrofrequencies. 
These events are categorized into three types based on their harmonic properties: (1) fundamental-only (no higher harmonics), (2) electrostatic (ES) harmonics, and (3) electromagnetic (EM) harmonics.
In this section, we first present three representative EMIC events observed by Van Allen Probe B (VAP-B), and then provide the statistical results based on the Van Allen Probes dataset.

\begin{figure}  
	\centering  
	\includegraphics[width=1.2\textwidth]{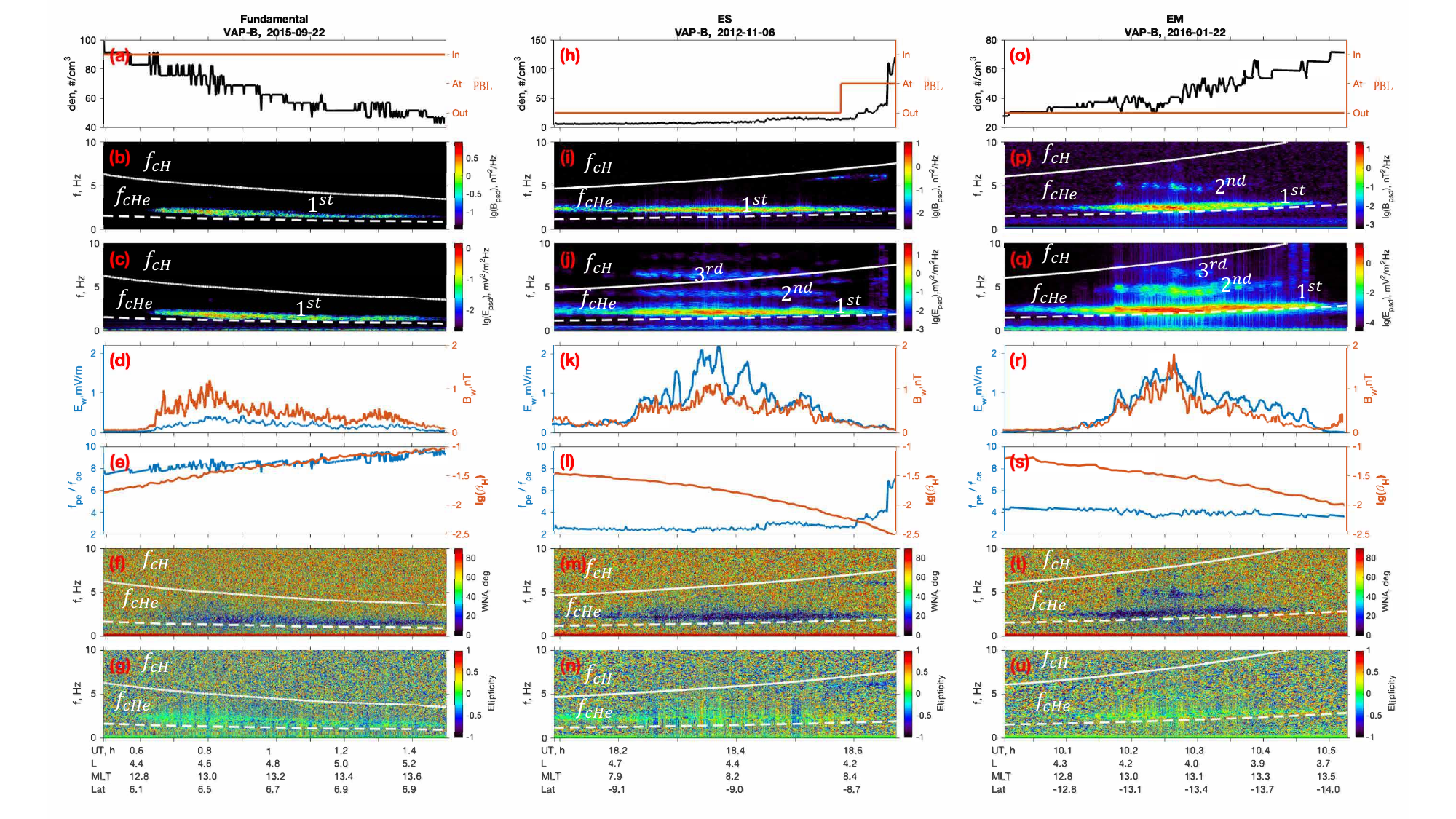}  
	\caption{ Three types of EMIC waves with only fundamental waves, electromagnetic (EM) harmonics, and electrostatic (ES) harmonics. (a, h, o): Background plasma density (in black) and satellite relative position to the PBL (in red). (b, i, p): Wave electric field power spectrum density. (c, j, q): Wave magnetic field power spectrum density. (d, k, r): Wave  electric (in blue) and magnetic (in red) field magnitude. (e, l, s): $f_{pe}/f_{ce}$ (in blue) and background plasma proton $\beta$ (in red). (f, m, t): Wave normal angle. (g, n, u): Ellipticity. The white solid and dashed lines in subfigures represent the gyrofrequency of proton $f_{cH+}$ and helium ion $ f_{cHe+}$, respectively. }
	\label{events}
\end{figure}

Figure \ref{events} shows examples of the three types of harmonic EMIC waves observed by Van Allen Probe (VAP-B).
The left column (Fig \ref{events}a-\ref{events}g) shows a fundamental harmonic EMIC wave event on September 22, 2015. 
Figure \ref{events}a shows the observed plasma density (black) and relative relation to the PBL (red).
The electric and magnetic field power spectral densities (PSDs) are presented in Figure \ref{events}b and \ref{events}c, respectively. A single narrow-band EMIC wave is observed with frequency close to the helium ion gyrofrequency $f_{cHe+}$, occurred inside the plasmaphere. 
The bandwidth $\delta f$ is less than 0.5 Hz, and the ratio of the bandwidth to the proton gyrofrequency $\delta f/ f_{cH+}$ is smaller than 0.1.
This type of EMIC wave is classified as a fundamental harmonic EMIC wave. 
In Figure \ref{events}d, we show the electric (blue) and magnetic (red) wave amplitudes. 
Figure \ref{events}e presents the ratio of the plasma frequency to the electron cyclotron frequency $f_{pe}/f_{ce}$ with a value close to $10$ (blue), and background proton beta $\beta_H$ (red). They both slightly increase in this period.
The WNAs are shown in Figure \ref{events}f, which indicates that these waves are nearly parallelly propagating. 
The wave polarization is primarily linear, as shown in Figure \ref{events}g.
The middle column (Figure \ref{events}h-\ref{events}n) show an ES harmonic EMIC wave event on November 16, 2012.
The wave event is observed outside the plasmasphere during the VAP-B inbound pass at $\sim$ 18.2-18.7 UT (Figure \ref{events}h). 
Compared to the fundamental case (Figure \ref{events}b and \ref{events}c), besides the fundamental mode, the electric field PSDs in Figure \ref{events}j have the second and third harmonic emissions.
Since the high harmonics have no corresponding magnetic fluctuation, we classify this type of EMIC waves as an ES harmonic EMIC wave. 
The electric amplitude (Figure \ref{events}k) of this ES harmonic event (around 2 mV/m) is much larger than that (Figure \ref{events}d) of the fundamental event (around 0.2 mV/m).
The $f_{pe} / f_{ce}$ of the ES harmonic EMIC wave event (around 2, Figure \ref{events}l) is much smaller than that of the fundamental event (around 10, Figure \ref{events}e). 
Similar to the fundamental case, Figure \ref{events}m and \ref{events}n show the WNAs and wave ellipticity.
In Figure \ref{events}o-\ref{events}u, we show an EM harmonic EMIC wave event observed on January 22, 2016, which occurred outside the plasmasphere.
Compared with the electric and magnetic PSDs of the fundamental and ES harmonic events (Figure \ref{events}b\&c,i\&j), the second harmonic exists in the magnetic field. 
We name this type of EMIC event as electromagnetic (EM) harmonic EMIC waves. 
Compared with the fundamental case in Figure \ref{events}d, the electric and magnetic field amplitudes are both larger in the EM harmonic case (Figure \ref{events}r). 

\begin{figure}  
	\centering  
	\includegraphics[width=1.0\textwidth]{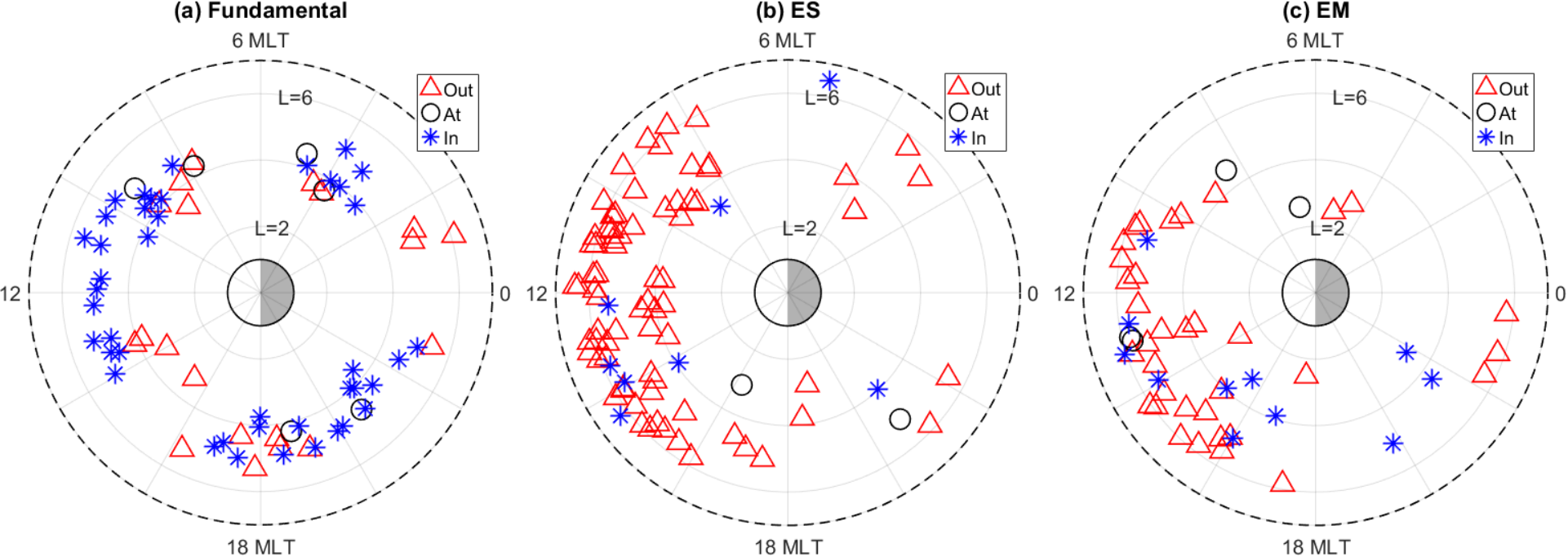}  
	\caption{Event distributions across $L$-shell/MLT of: (a) Fundamental, (b) ES harmonic, and (c) EM harmonic EMIC events. The blue asterisks represent events inside the plasmasphere, the red triangles represent the events outside the plasmasphere, and the black circles represent the events at the PBL. }
	\label{mlt}
\end{figure}

We perform a statistical study to address the statistical characteristics of these three types of EMIC waves.
In total, we identify 72 fundamental EMIC wave events, 85 ES harmonic EMIC events, and 53 EM harmonic EMIC wave events.
For each event, we record the time corresponding to the maximum electric field amplitude. 
The distribution of events across $L$-shell and magnetic local time (MLT) is shown in Figure \ref{mlt}, where the terms “In”, “Out”, and “At” indicate the locations inside the plasmasphere, outside the plasmasphere, and at the PBL, respectively, as mentioned in the section 2.
Figure \ref{mlt}a shows that over 60$\%$ of the fundamental EMIC waves occur inside the plasmasphere (blue asterisks) or at the PBL region (black circles), with no explicit dependency on the MLT. 
In contrast, Figure \ref{mlt}b and \ref{mlt}c demonstrate that more than 80$\%$ ES harmonic EMIC waves and nearly 70$\%$ EM harmonic EMIC waves happen outside the plasmasphere, with nearly 90$\%$ ES harmonic EMIC waves and 85$\%$ EM harmonic EMIC waves occurring on the dayside. 

\begin{figure}  
	\centering  
	\includegraphics[width=1.0\textwidth]{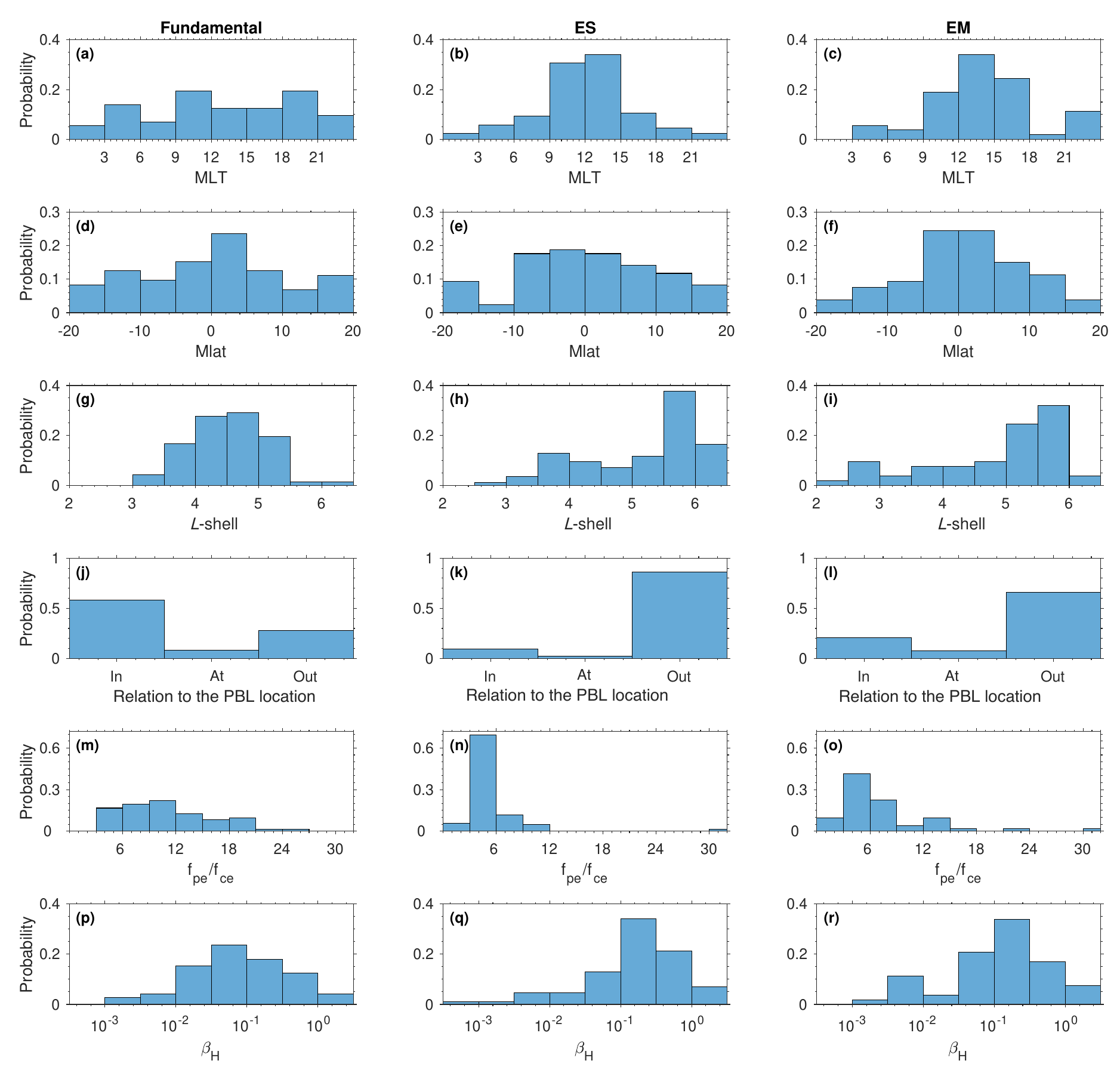}  
	\caption{Statistical results of the position and plasma conditions of three types of EMIC waves. (a-c) Magnetic Local Time (MLT);(d-e) Magnetic latitude (Mlat); (g-i) $L$-shell; (j-l) The relative position to the PBL; (m-o) $f_{pe}/f_{ce}$  ; (p-r) Proton beta $\beta_H$. }
	\label{fig_bg}
\end{figure}

In Figure \ref{fig_bg}, we present the dependencies of EMIC events on the position and background plasma conditions. 
The probability in each bin is computed as the ratio of the number of events in each bin to the total number of events for each type.
Figure \ref{fig_bg}a-\ref{fig_bg}l illustrate the probability distribution of events based on their positions, including MLT, magnetic latitude (Mlat), $L$-shell, and their relation to PBL location. 
Figure \ref{fig_bg}a-\ref{fig_bg}c show that both EM and ES harmonic EMIC waves are more concentrated on the dayside, mainly at 9-15 MLT, compared to the fundamental EMIC waves. 
In Figure \ref{fig_bg}d-\ref{fig_bg}f, we find that these three types of events are most frequently observed near the geomagnetic equator. 
This trend is consistent with the excitation of EMIC waves near the equator due to the injection of anisotropic energetic ring current ions \cite{ChenLJ_2010JGR_emic, Jordanova_2008JGR}. Consequently, their harmonics, when present, are also expected to be predominantly located in the same region. Notably, EM harmonic EMIC waves occur more frequently near the equator compared to ES harmonic EMIC waves.
Figure \ref{fig_bg}g shows that nearly $80 \%$ of fundamental EMIC events occur at $L<5$, and $60 \%$ of these events take place inside the plasmasphere (Figure \ref{fig_bg}j). 
Conversely, more than $60 \%$ of ES and EM harmonic events are observed at $L>5$ (Figure \ref{fig_bg}h,i), and over 70$\%$ occur outside the plasmasphere (Figure \ref{fig_bg}k,l).
Figure \ref{fig_bg}m-o and \ref{fig_bg}p-r show the dependencies on $f_{pe}/f_{ce}$ and proton beta $\beta_H$. 
Compared with the fundamental events, the harmonic cases tend to occur with smaller $f_{pe}/f_{ce}$ ($<$ 9) and larger proton $\beta_H$ ($>$ 0.1). 
The mean values of these parameters of the three types of harmonic EMIC waves are summarized in Table \ref{table}. 
It shows the fundamental events have smaller L$\sim$4.5 and larger ${f_{pe}/f_{ce}} \sim$10, while the harmonic events have larger L$\sim$5 and smaller ${f_{pe}/f_{ce}} \sim$6.

\begin{figure}  
	\centering  
	\includegraphics[width=1.0\textwidth]{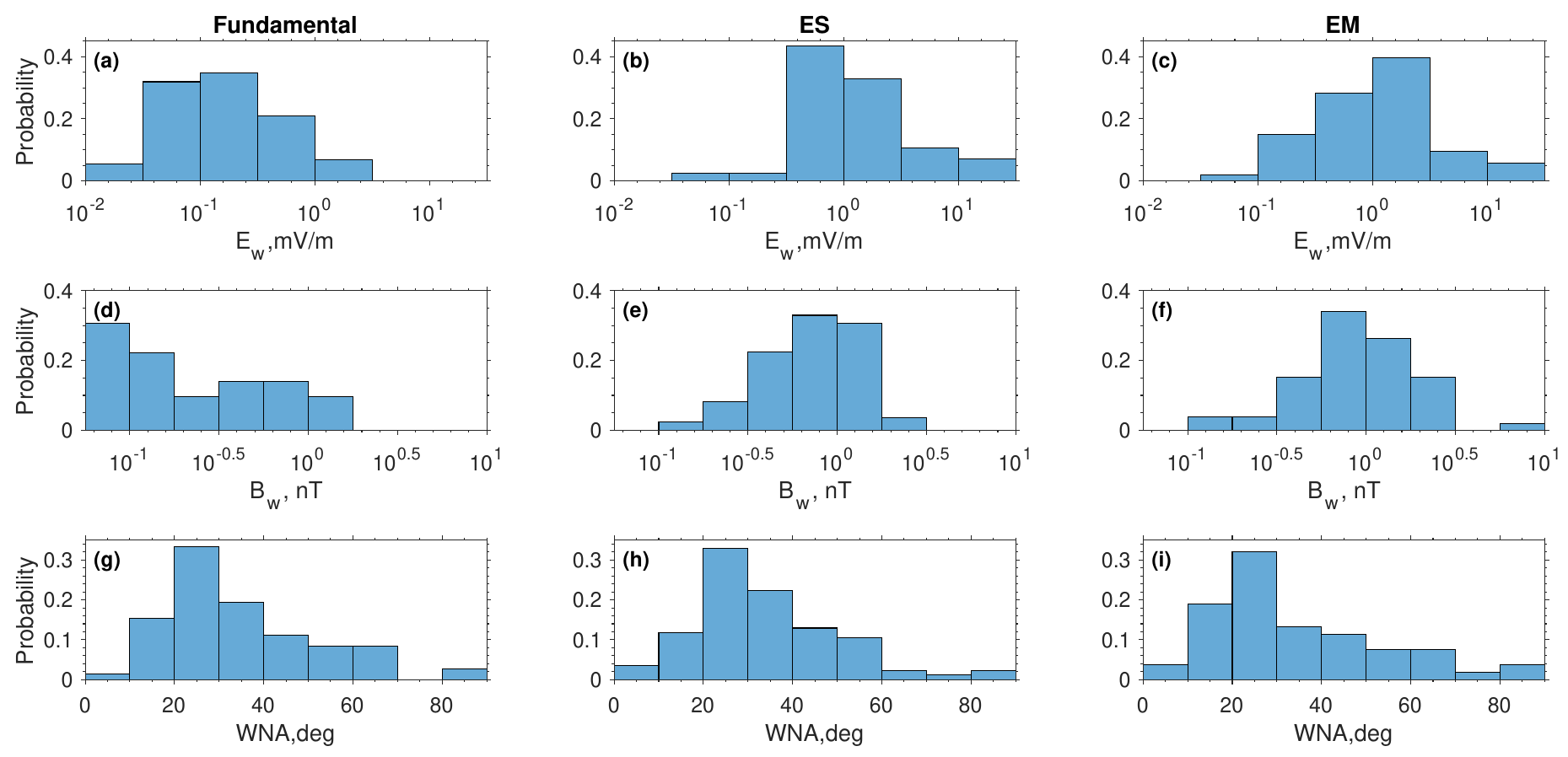}  
	\caption{ Statistical results of the wave properties of the fundamental mode of the three types of EMIC waves:  (a-c) Wave electric field amplitude, (d-f) Wave magnetic field magnitude, and (g-i) Wave normal angle. }
	\label{fig_wave}
\end{figure}

Figure \ref{fig_wave} summarizes the event dependencies of wave properties.
Panels a-c and d-f display the probability distributions of wave electric field amplitude $E_w$ and magnetic field amplitude $B_w$, respectively. 
Note that, even though the wave electric and magnetic amplitudes of EMIC harmonics are considerably large, they are still several orders of magnitude smaller than those of the fundamental modes. Thus, The wave electric and magnetic amplitudes here are mainly contributed from the fundamental mode.
More than $85 \%$ ES and EM events are observed with $E_w > 0.32$ mV/m and $B_w > 0.32$ nT, whereas nearly $70 \%$ fundamental events happen with $E_w < 0.32$ mV/m and $B_w < 0.32$ nT. 
This indicates that both ES and EM harmonic EMIC events are associated with a stronger elctromagnetic fundamental mode.
Figure \ref{fig_wave}g-\ref{fig_wave}i show the probability distributions of WNAs, revealing no significant differences in the WNA distributions among the three types of EMIC waves.
In Table \ref{table}, the mean values of $E_w$ and $B_w$ support the conclusions drawn from the statistical results. 
Specifically, for ES harmonic events, $E_w$ is nearly 10 times greater than that in the fundamental events, while $B_w$ is nearly 2 times greater. For EM harmonic events, $E_w$ is nearly 8 times greater, and $B_w$ nearly 3 times greater compared to the fundamental events.
The stronger fundamental mode in the harmonic event is aligned with previous studies \cite{ZhuH_2019GRL, DengD_2022GRL}, indicating the nonlinear wave-wave excitation of the harmonic EMIC waves, which requires large amplitudes of fundamental waves.

\begin{table}[p]
	\caption{Mean value of the parameters in Figure \ref{fig_bg} and \ref{fig_wave}. }
	\begin{center}
		\begin{tabular}{| c | c | c | c |} \hline
			\textbf{Parameters $\backslash$ Categories} & \textbf{Fundamental } & \textbf{ES} & \textbf{EM} \\  \hline
			\textbf{MLT(h)} & 12.96  & 12.26 & 13.96 \\  \hline
			\textbf{MLat(deg)} & 0.23  & 0.68 & 0.83 \\  \hline
			\textbf{Lshell} & 4.5 & 5.18  & 4.86   \\  \hline
			$\mathbf{f_{pe}/f_{ce}}$ & 10.00   & 6.03  & 5.89  \\  \hline
			$\mathbf{log_{10}(\beta_{H})}$ &  -1.09  & -0.83  & -0.93  \\  \hline
			$\mathbf{E_w(mV/m)}$ & 0.31  & 2.93  &  2.56  \\  \hline
			$\mathbf{B_w(nT)}$ & 0.37   & 0.83  & 1.18  \\  \hline
			\textbf{WNA(deg)} & 34.51 & 34.92 & 33.6 \\  \hline
		\end{tabular}
	\end{center}
	\label{table}
\end{table}

\section{Conclusions and Discussion}
In this paper, we present, for the first time, a statistical study of the harmonic EMIC waves based on observations from the Van Allen Probes between 2012 and 2019. We focus on narrow-band EMIC waves, and categorize them into three groups based on the presence of harmonics. We then compare the properties of these three types of EMIC waves. The main conclusions are summarized as follows:

(1) Harmonic EMIC waves are predominantly observed outside the dayside plasmasphere with $L >5$.

(2) Harmonic EMIC waves tend to occur in regions with a low $f_{pe}/f_{ce}$ ($<9$) and a high proton $\beta_H$ ($>0.1$). 

(3) Harmonic EMIC waves are mostly associated with strong fundamental modes ($E_w > 0.32$ mV/m and $B_w > 0.32$ nT), indicating nonlinear processes in the harmonic excitation.

For now, the statistical differences in wave properties between ES and EM cases are not prominent. 
The underlying mechanisms responsible for these two types of harmonic EMIC waves remain controversial. 
Some works suggested that nonlinear wave-wave coupling is responsible for exciting both two types of harmonic waves \cite<e.g.,>{ZhuH_2019GRL, DengD_2022GRL}. 
\citeA{Sauer_2022JGR} proposed that, in a multi-species plasma, energy and momentum exchange between different ion components could explain the origin of ES harmonic EMIC waves. 
Thus, identifying the distinct generation mechanisms behind these two types of waves will be the focus of our future work. 

EMIC waves play a vital role in scattering electrons in the radiation belt, with previous studies primarily focusing on the interaction between electrons and monochromatic EMIC waves.  
\citeA{AnX_2014JGR} proposed a two-wave model, showing that interactions with two-wave EMIC waves can induce more stochastic scattering of MeV electrons compared to single-wave interactions, as demonstrated through theory and test-particle simulations. In the future, we plan to investigate the response of electrons to harmonic EMIC waves, and compare this with their response to monochromatic EMIC waves based on observations.

This study confirms the existence of narrow-band intense EMIC waves. The higher harmonic components, when present, are significantly weaker than the fundamental component, but may still influence the saturation of the fundamental wave amplitude. 
Future research, using particle-in-cell simulations, will aim to explore the underlying nonlinear physics involved in these processes.

\section*{Open Research Section}
The data of Van Allen Probes used in this paper are provided by the Space Physics Data Facility (SPDF) (https://spdf.gsfc.nasa.gov/pub/data/rbsp/). 
The EMIC wave event list used in this paper is in the dataset \cite{Gu_2024zenodo}. 

\acknowledgments
We acknowledge the support of NASA Grant 80NSSC19K0283, 80NSSC21K0518.

%
%

\bibliography{ ref.bib }

\end{document}